\documentclass[prl,twocolumn,superscriptaddress,showpacs,floatfix,amsfonts]{revtex4}
\usepackage{graphicx,graphics,color,epsfig}
\usepackage{bm}
\usepackage{amsmath}
\usepackage{amssymb}
\begin{document}
\preprint{}
\title{Effects of
a Collective Spin Resonance
Mode on
the STM
Spectra of D-Wave Superconductors}
\author{Jian-Xin Zhu}
\affiliation{Theoretical Division, MS B262, Los Alamos National
Laboratory, Los Alamos, New Mexico 87545}
\author{Jun Sun}
\affiliation{Department of Physics \& Astronomy, Rice University,
Houston, Texas 77005}
\author{Qimiao Si}
\affiliation{Department of Physics \& Astronomy, Rice University,
Houston, Texas 77005}
\author{A. V. Balatsky}
\affiliation{Theoretical Division, MS B262, Los Alamos National
Laboratory, Los Alamos, New Mexico 87545}

\begin{abstract}
A high-energy spin resonance mode
is known to exist in many high-temperature
superconductors. Motivated by recent scanning tunneling microscopy
(STM) experiments in superconducting
Bi$_2$Sr$_2$CaCu$_2$O$_{8+\delta}$, we study the effects of this
resonance mode on the local density of states (LDOS). The coupling
between the electrons in a $d$-wave superconductor and the
resonance mode produces high-energy peaks in the LDOS, which
displays a two-unit-cell periodic modulation around a nonmagnetic
impurity. This suggests a new means to not only detect the
dynamical spin collective mode but also study its coupling to
electronic excitations.

\end{abstract}
 \maketitle

A prominent feature in the excitation spectrum of the high-$T_c$
superconductors is the ``41 meV'' collective spin resonance mode,
seen by inelastic neutron scattering experiments in most of the
cuprate families~\cite{Rossat-Mignod91,Mook93,Fong96,He01}. The
physics of this resonance mode -- including its microscopic
origin, its connections with other physical properties, as well as
its role on superconductivity itself -- has been the subject of
considerable debate. Given the recent developments of the atomic
resolution scanning tunneling microscopy
(STM)~\cite{Hoffman02a,Hoffman02b,Howald02a,Howald02b,McElroy02},
it is timely to address the possible manifestation of this
resonance mode in the local density of states (LDOS). A number of
theoretical works~\cite{Polkovnikov02,Podolsky02,Han02,Kivelson02}
have addressed the effect of related spin fluctuations on the
LDOS. These works focused on the pinning of the spin fluctuations
by impurity: a dynamical spin mode centered around the wavevector
${\bf Q}$ leads to a $2{\bf Q}$ spatial modulation in the
low-energy LDOS. This result is smoothly connected to what happens
in the case of a static spin-density-wave
ordering~\cite{Zachar98,Zhu02}. However, for the resonance mode --
which is sharply peaked at ${\bf Q} = ({\pi,\pi})$ -- such effects
would not be manifested [since $2 {\bf Q} = ({2\pi,2\pi})$ is
equivalent to $(0,0)$]. There are also works about quasiparticle
scattering from nomagnetic impurities~\cite{Wang02,Zhang02}.

In this paper, we show that the coupling of $d$-wave
quasiparticles to the resonance mode does produce spatial
modulations in the LDOS around an impurity. The feature is located
at relatively high energies, $\approx \pm (\Delta_0 + \Omega_0)$,
where $\Delta_0$ is the maximum superconducting energy gap and
$\Omega_0$ the resonance energy. In addition, the wavevector of
the LDOS modulation is close to ${\bf Q} =(\pi,\pi)$. Our
predictions
could be observable
by the STM experiments. Such
STM studies represent a new means to characterize the coupling
between the electronic excitations and the resonance mode. The STM
feature we discuss relates to the ``peak-dip-hump'' structure of
the angle resolved photoemission spectroscopy
(ARPES)~\cite{Dessau91,Ding96,Abanov02,Kee02,Eschrig00,Abanov00};
the inference about
the electron-spin coupling from the ARPES and related
spectroscopies is a topic of recent
controversy~\cite{Kee02,Abanov02} and we hope that the STM studies
we propose will shed new light on this important issue.

We start with a model Hamiltonian describing two-dimensional
electrons coupled to a collective spin mode and in the presence of
a single-site  impurity: $\mathcal{H}=\mathcal{H}_{BCS}+
\mathcal{H}_{sp} + \mathcal{H}_{imp}$. Here the BCS-type
Hamiltonian for a uniform $d$-wave superconductor is given by $
\mathcal{H}_{BCS}=\sum_{\mathbf{k},\sigma}
(\varepsilon_{\mathbf{k}}-\mu) c_{\mathbf{k}\sigma}^{\dagger}
c_{\mathbf{k}\sigma} +\sum_{\mathbf{k}}(\Delta_{\mathbf{k}}
c_{\mathbf{k}\uparrow}^{\dagger}c_{-\mathbf{k}\downarrow}^{\dagger}
+\Delta_{\mathbf{k}}^{*}
c_{-\mathbf{k}\downarrow}c_{\mathbf{k}\uparrow}) $, where
$c_{\mathbf{k}\sigma}^{\dagger}$ ($c_{\mathbf{k}\sigma}$) creates
(annihilates) a conduction electron of spin $\sigma$ and
wavevector $\mathbf{k}$, $\varepsilon_{\mathbf{k}}$  is the normal
state energy dispersion for the conduction electrons, $\mu$ the
chemical potential, and
$\Delta_{\mathbf{k}}=\frac{\Delta_{0}}{2}(\cos k_x -\cos k_y)$ the
$d$-wave superconducting energy gap. The coupling between the
electrons and the resonance mode is modeled by an interaction term
$\mathcal{H}_{sp}=g\sum_{i} \mathbf{S}_{i}\cdot \mathbf{s}_{i}$,
where the quantities $g$, $\mathbf{s}_{i}$, and  $\mathbf{S}_{i}$
are the coupling strength, the electron spin operator at site $i$,
and the operator for the collective spin degrees of freedom,
respectively. The dynamics of the collective mode will be
specified below. The impurity scattering is given by
$\mathcal{H}_{imp}=U_{0}\sum_{\sigma} c_{0\sigma}^{\dagger}
c_{0\sigma}$, where without loss of generality we have taken a
single-site impurity of strength $U_0$ located at the origin,
$\mathbf{r}_{i}=0$. By introducing a two-component Nambu spinor
operator, $\Psi_{\mathbf{k}}=(c_{\mathbf{k}\uparrow},
c_{-\mathbf{k}\downarrow}^{\dagger})^{T}$, the matrix Green's
function for the $d$-wave BCS Hamiltonian $\mathcal{H}_{BCS}$ is
determined by $
G_{0}^{-1}(\mathbf{k};i\omega_{n})=\left( \begin{array}{cc}
i\omega_{n}-\xi_{\mathbf{k}} & -\Delta_{\mathbf{k}} \\
-\Delta_{\mathbf{k}} & i\omega_{n}+\xi_{\mathbf{k}}
\end{array}
\right)\;,
$
where $\xi_{\mathbf{k}}=\varepsilon_{\mathbf{k}}-\mu$
and $\omega_{n}=(2n+1)\pi T$ is the fermionic
Matsubra frequency.
We have also assumed that the $d$-wave pair
potential is real.

\begin{figure}[th]
\centerline{\psfig{file=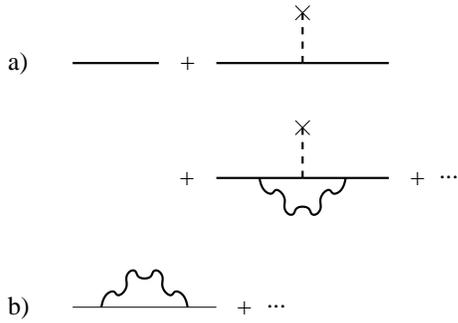,width=6cm}} 
\caption{a) Diagrams for ${G}({\bf p},{\bf p+q};\omega)$.
The thick solid line represents the full conduction electron
Green's function in the absence of impurity,
$\underline{{G}}_{0}$.
The dashed line is the
coupling to an impurity, specified by a cross. The wavy
line denotes the propagator of the collective mode;
b) Self-energy diagrams for the conduction electrons in the absence
of impurity, $\hat{\Sigma}$.
The thin solid line is the bare (BCS) conduction
electron Green's function, ${{G}}_{0}$. }
\label{FIG:diagrams}
\end{figure}

For a homogeneous system, where only the
inelastic scattering of quasiparticles from the collective mode
occurs, we calculate the self-energy to the second order in the
coupling constant (see Fig.~\ref{FIG:diagrams}b):
$$
\hat{\Sigma}(\mathbf{k};i\omega_{n})=\frac{3g^{2}T}{4}\sum_{\mathbf{q}}
\sum_{\Omega_{l}} \chi(\mathbf{q};i\Omega_{l})
G_{0}(\mathbf{k}-\mathbf{q};i\omega_{n}-i\Omega_{l})\;,
$$
where $\chi(\mathbf{q};i\Omega_{l})$ is the dynamical spin
susceptibility $\chi_{ij}(\tau)=\langle T_{\tau}
(S^{x}_{i}(\tau)S^{x}_{j}(0))\rangle$ and
 $\Omega_{l}=2l\pi T$ the bosonic
Matsubra frequency.
The dressed Green's function is:
\begin{equation}
\underline{{G}}_{0}^{-1}(\mathbf{k};i\omega_{n})=\left(
\begin{array}{cc}
i\omega_{n}-\xi_{\mathbf{k}}-\Sigma_{11} & -\Delta_{\mathbf{k}}-\Sigma_{12} \\
-\Delta_{\mathbf{k}}-\Sigma_{21} &
i\omega_{n}+\xi_{\mathbf{k}}-\Sigma_{22}
\end{array}
\right)\;.
\end{equation}
The corresponding real-space dressed Green's function
$\underline{{G}}_{0}(i,j;i\omega_{n})$ is obtained through
a Fourier transform with respect to
$\mathbf{r}_{i}-\mathbf{r}_{j}$. For the $d$-wave pairing
symmetry, one can show that the local Green's function,
$\underline{\hat{g}}_{0}(i\omega_{n})=
\underline{{G}}_{0}(i,i;i\omega_{n})$
is diagonal. In the presence of a single-site impurity at
$\mathbf{r}_{i}=0$ with potential strength $U_0$,
the site-dependent Green's function
can be written in terms of the $T$-matrix:
\begin{eqnarray}
{G}(i,j;E)&=&{\underline{G}}_{0}(i,j;E)\nonumber \\
&&+\sum_{lm} {\underline{G}}_{0}(i,l;E)\hat{T}_{lm}(E)
{\underline{G}}_{0}(m,j;E)\;.
 \label{gij}
\end{eqnarray}
Due to the vertex corrections induced by the coupling to
the collective modes (Fig.~\ref{FIG:diagrams}a),
the $T$ matrix in general contains
site-off-diagonal terms.
We will first carry out the calculation without
the vertex corrections, in which case
$\hat{T}_{lm} = \hat{T} \delta_{l,0} \delta_{m,0}$,
with
$\hat{T}^{-1}=U_{0}^{-1} \sigma_{3}-\hat{\underline{g}}_0 $,
where $\sigma_{3}$ is the
$z$-component of the Pauli matrix. The LDOS at the $i$-th site,
summed over two spin components,
is
\begin{eqnarray}
\rho(\mathbf{r}_{i},E)=-\frac{2}{\pi} \mbox{Im} G_{11}(i,i;E+i\gamma)\;,
\label{ldos}
\end{eqnarray}
where $\gamma=0^+$.

Up to now, our discussion and formulation are quite general
and can be used to study the effects of any
dynamic mode once the susceptibility $\chi$ is known.
We treat the susceptibility in a phenomenological
form (based on the
inelastic neutron scattering observations),
see also~\cite{Eschrig00}:
\begin{equation}
\chi(\mathbf{q};i\Omega_{l})=-\frac{\delta_{\mathbf{q},\mathbf{Q}}}{2}
\left[\frac{1}{i\Omega_{l}-\Omega_{0}}-\frac{1}
{i\Omega_{l}+\Omega_{0}}\right]\;,
\end{equation}
where we denote the wavevector $\mathbf{Q}=(\pi,\pi)$ and the mode
energy by $\Omega_{0}$. This form is especially suitable for the
optimally doped YBa$_2$Cu$_3$O$_{6+y}$ (YBCO) compounds in the
superconducting phase, where the observed neutron resonance peak
is
almost resolution-limited in energy and fairly sharp in
wavevector.
The resonance peak in BSCCO is
broadened in both energy and wavevector.
The finite width in the wavevector space
might be important for the
ARPES lineshape in general
and in particular
the understanding of
the ARPES spectra
away from the $M$ points [$\mathbf{k}=(\pi,0)$
and symmetry-related points] of the
Brillouin zone~\cite{Eschrig00}, but should not
change the
qualitative
conclusion of our work: the LDOS effects we will
discuss arise from the fact that the dominant effects
of the resonance mode on the single-electron spectral
functions occur near the $M$ points which is expected to
remain to be the case beyond our simplified
form for the susceptibility.
In addition,
given that
the peak in BSCCO is still quite sharp in energy,
we expect that the main effect of the broadening in energy
of the resonance mode is to extend the bias window
for the LDOS feature we will discuss.
We have also neglected the incommensurate peaks
seen in the inelastic
neutron scattering experiments
in YBCO (the part that disperses
``downward'' away from the resonance
peak)~\cite{Arai99,Fong00,Brinckmann99,Kao00}, since their
spectral weight is significantly smaller than that of the
resonance mode. For the normal-state energy dispersion, we use
$\varepsilon_{\mathbf{k}}=-2t (\cos k_x + \cos k_y) -4t^{\prime}
\cos k_{x} \cos k_y$, where $t$ and $t^{\prime}$ are the nearest
and next-nearest neighbor hopping integral. Unless specified
explicitly, the energy is measured in units of $t$. We choose
$t^{\prime}=-0.2$ to model the band structure of the hole-doped
cuprates.  Since the maximum energy gap for most of the cuprate
superconductors at the optimal doping is about $30 \;\mbox{meV}$
while the resonance mode energy is in the range between $35$ and
$47$ meV, we take $\Delta_0=0.1$ and $\Omega_0=0.15$ (i.e., $1.5
\Delta_{0}$).  The chemical potential is tuned to give an optimal
doping value 0.16. To mimic the intrinsic life time broadening, in
our numerical calculation we take $\gamma$ of Eq.~(\ref{ldos}) to
be $0.04\Delta_{0}$. A system size of $N_x\times N_y=1000 \times
1000$ is taken in the numerical calculation.

\begin{figure}[th]
\centerline{\psfig{file=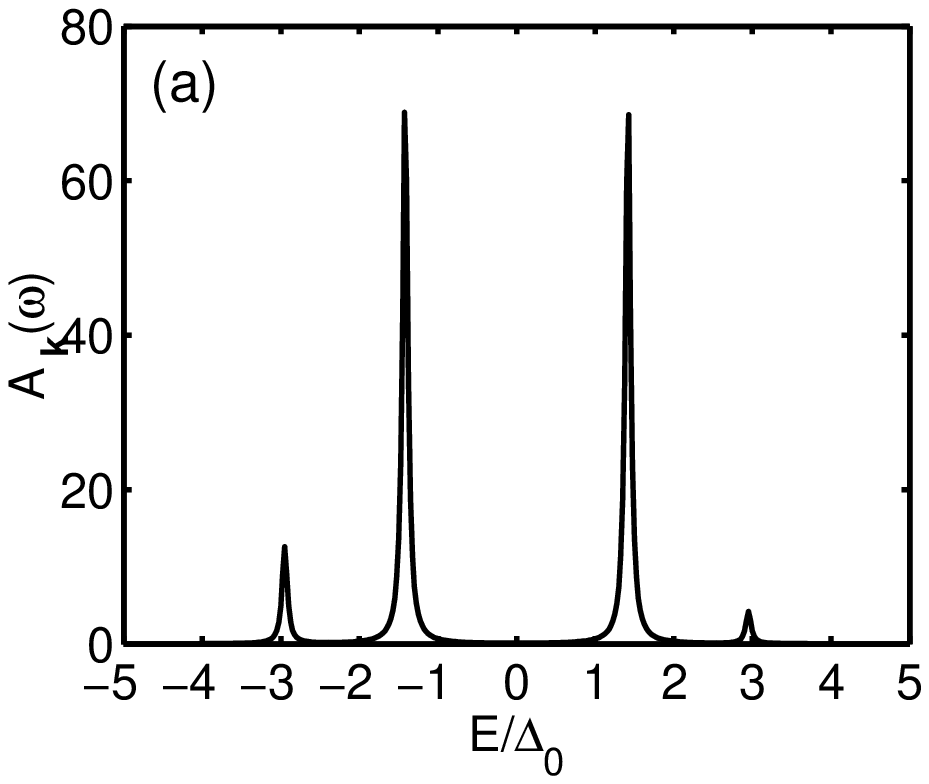,width=4.3cm}
\psfig{file=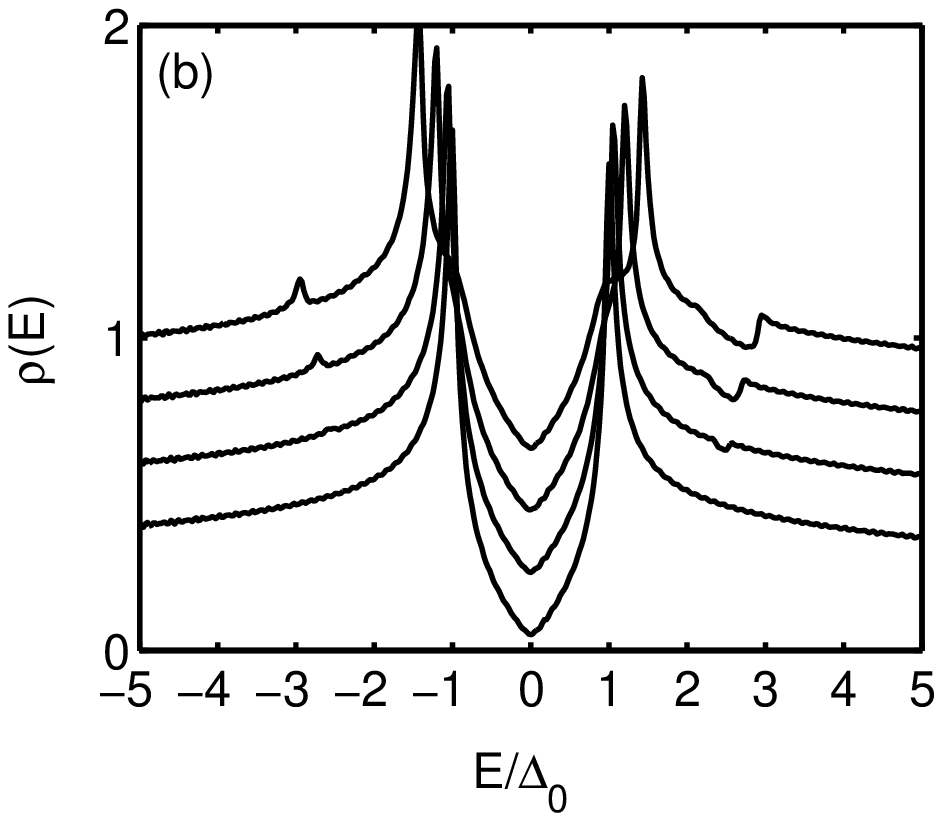,width=4.3cm}} \caption{Spectral function
$A_{\mathbf{k}}(E)$ at the wavevector $\mathbf{k}=(\pi,0)$ for
$g/\Delta_{0}=3/\sqrt{2}$ (a).
 The density of states, $\sum_{\mathbf{k}}A_{\mathbf{k}}(E)$,
 is shown in (b), for
various values of $g/\Delta_{0}=0$, $1/\sqrt{2}$, $2/\sqrt{2}$,
and $3/\sqrt{2}$ (from lower to upper); here, for easier viewing,
the consecutive curves are shifted by 0.2 along the vertical axis.
} \label{FIG:BAND_DOS} \label{FIG:SPECTRUM}
\end{figure}

In the absence of impurities, the density of states is the
summation of the spectral function,
$A_{\mathbf{k}}(E)=-\frac{2}{\pi}
\mbox{Im}\underline{G}_{0,11}(\mathbf{k};E+i\gamma)$, over all
wavevectors $\mathbf{k}$. Fig.~\ref{FIG:SPECTRUM}(a) shows the
spectral function at an $M$ point of the Brillouin zone. Without
the electron-mode coupling, as is well known, the spectral
function is peaked at the maximum gap edges $\pm \Delta_0$. As the
electron-mode coupling is switched on, new peaks emerge at the
energies $\pm E_{1} \approx \pm (\Delta_{0}+\Omega_{0})$. (For
simplicity, we have neglected the broad ``background'' part of the
single-electron spectral function in our consideration.) The peaks
in $A_{\mathbf{k}}$ originate from the poles of the Green's
function $\underline{G}_{0,11}$. Note that the weight of the peak
at $-\Delta_0$ is larger than at $\Delta_0$ because the van Hove
singularity is below the Fermi energy. Since the spectral weight
of the spin resonance mode [i.e., $\mbox{Im}
\chi(\mathbf{q};\omega)$] is peaked at $\mathbf{Q}$, the feature
of the quasiparticle self-energy is the strongest around the $M$
points of the zone because they are connected by $\mathbf{Q}$. The
singularity in the quasiparticle self-energy causes additional
poles in the Green's function. As the coupling constant $g$
increases, these peaks are shifted to higher energies and, in
addition, their spectral weight is enhanced; simultaneously, the
weight of the superconducting coherent peaks is reduced to obey
the sum rule. The shift of states due to inelastic scattering is
expected in DOS and is also expected for scattering off local
mode~\cite{Balatsky02}. Fig.~\ref{FIG:BAND_DOS}(b) plots the
density of states and clearly shows that the high energy peaks
still occur around $\pm E_{1}$. In other words, the contributions
from near the $M$ points dominate the wavevector summation for the
density of states, reflecting the flat nature of the normal state
band near this point. Furthermore, the highly asymmetrical
structure in the DOS at energies $-E_{1}$ and $E_{1}$ comes from
the singular structure in the quasiparticle self-energy. These
results, for the clean case, are consistent with earlier
studies of the ARPES~\cite{Dessau91,Ding96,Abanov02,Kee02,Eschrig00}
and DOS~\cite{Abanov00}.

\begin{figure}[th]
\centerline{\psfig{file=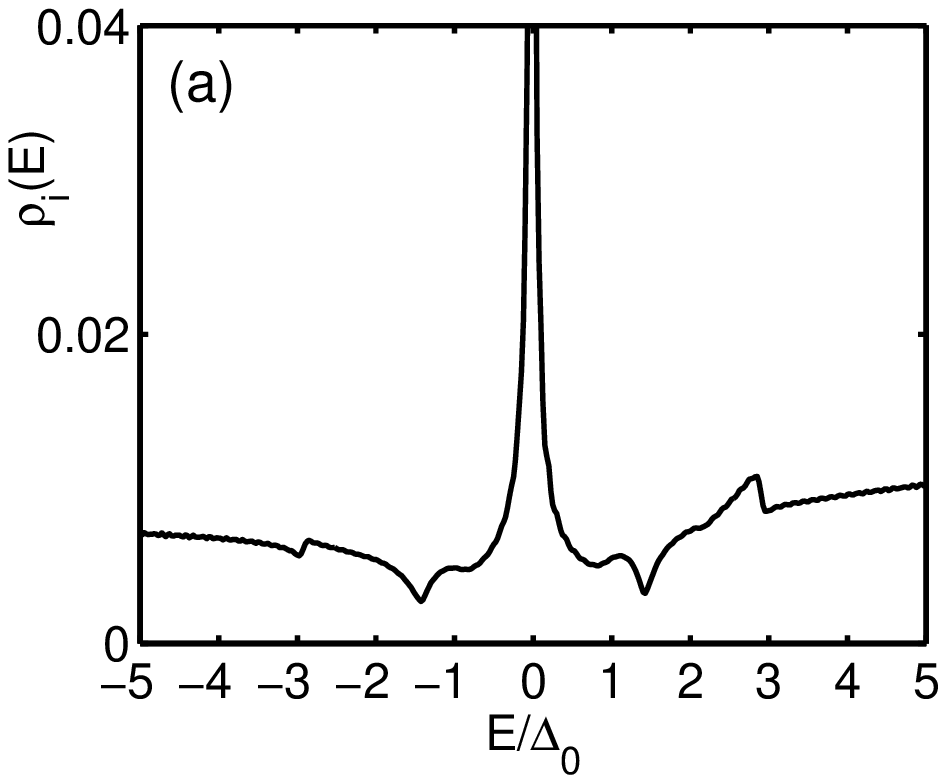,width=4.4cm}
\psfig{file=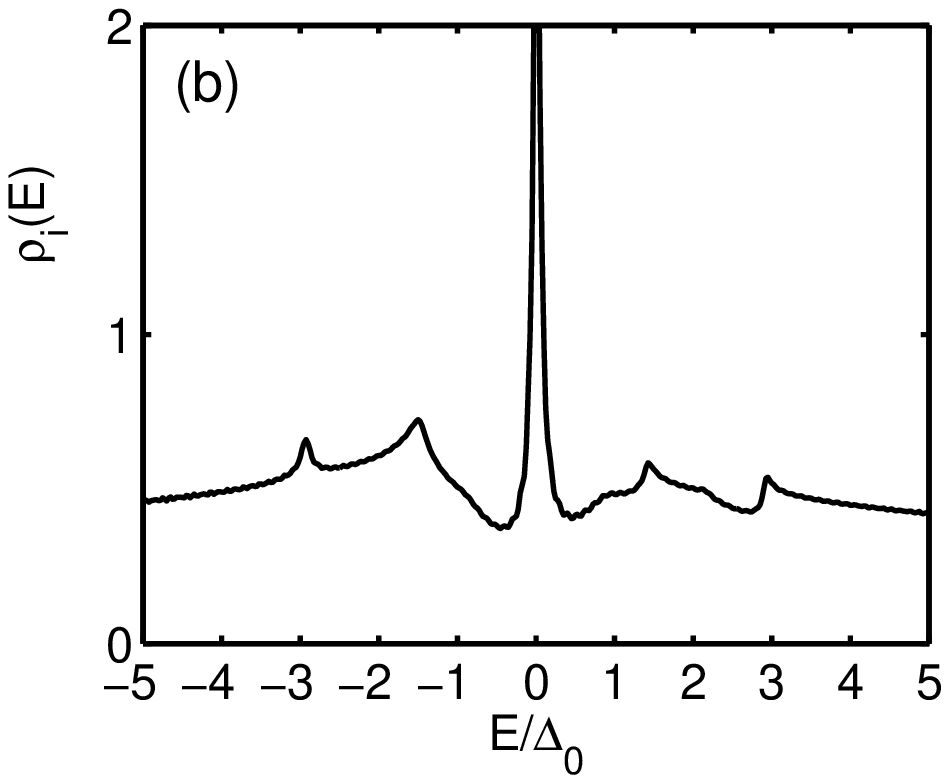,width=4.4cm}}
\caption{Local density of states at the impurity site (a) and at
its nearest neighbor (b). The coupling constant
$g/\Delta_{0}=3/\sqrt{2}$.} \label{FIG:LDOS}
\end{figure}

We are now in a position to address the LDOS in the presence of a single
nonmagnetic impurity. For concreteness, we take the on-site
potential $U_0=100\Delta_{0}$.
Fig.~\ref{FIG:LDOS} shows the LDOS directly at the impurity site,
as well as at its nearest
neighbor. The near-zero energy resonant state triggered by the
quasiparticle scattering from the impurity~\cite{Balatsky95} is
robust against the electron-mode coupling.
Our key new results are two-fold.
First, the impurity modifies the shape of the spectral features
at $\pm E_{1}$, which can now be either a dip or a peak.
Second, and more importantly, these high energy features
at $\pm E_{1}$ exhibit a spatial dependence.
At the impurity site, the LDOS displays a dip at $-E_{1}$ but
a peak at $+E_{1}$.
The behavior is reversed at the site closest to the impurity.

\begin{figure}[th]
\centerline{\psfig{file=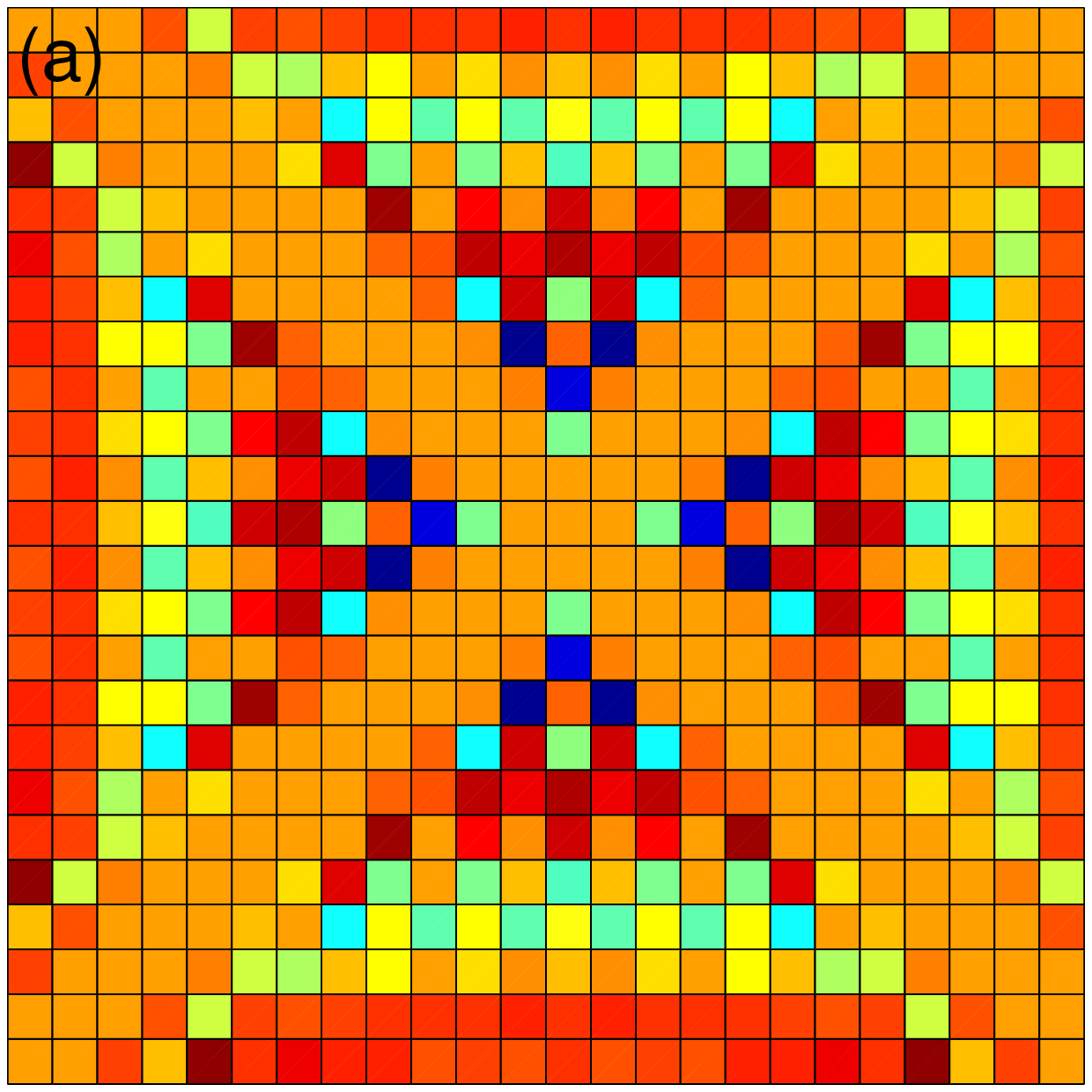,width=4.2cm}
\psfig{file=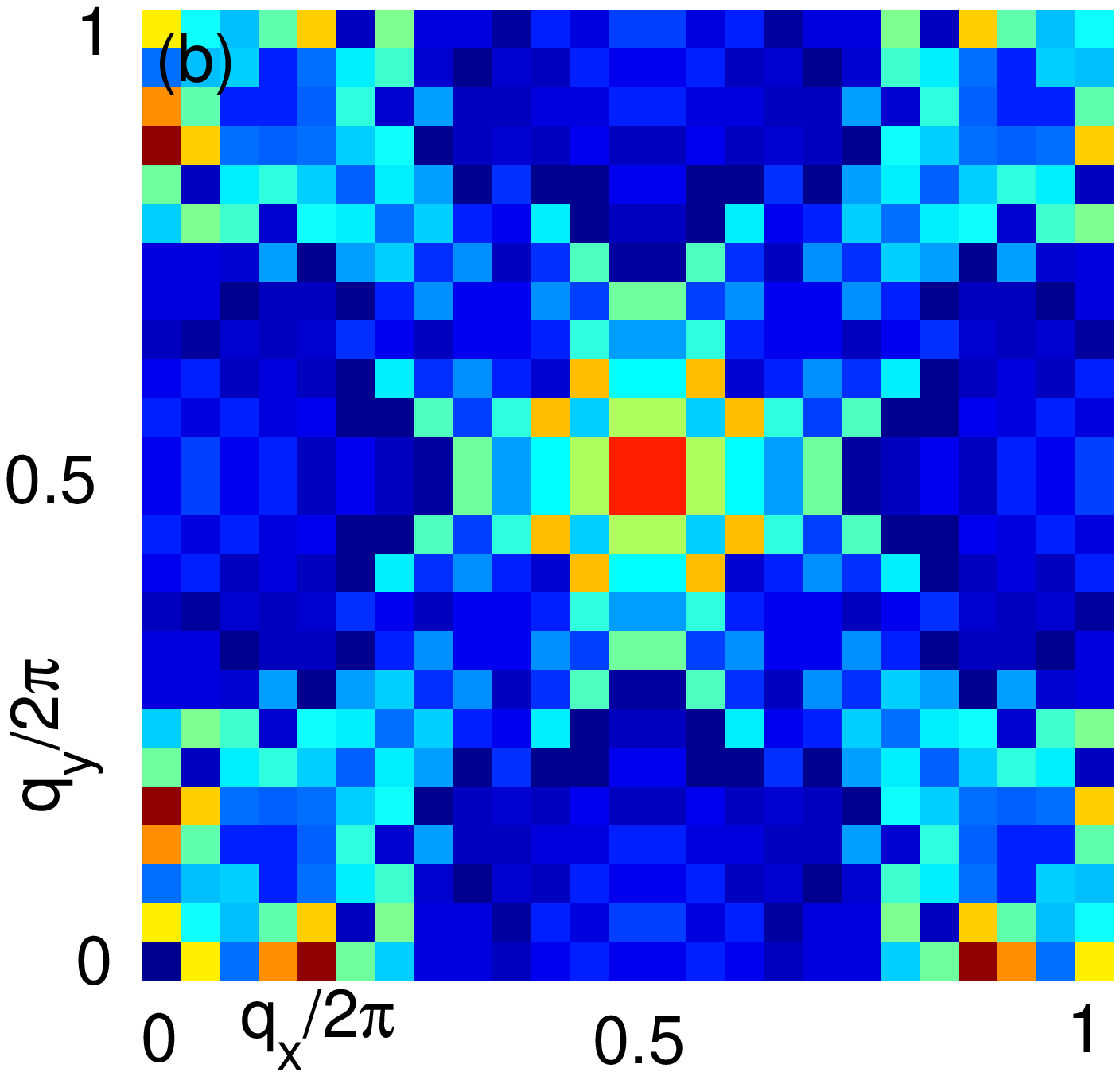,width=4.4cm}}
\caption{ The spatial variation of the LDOS around the impurity
with the mode coupling $g/\Delta_0=3/\sqrt{2}$ at $E=-E_{1}$ (a),
and its corresponding Fourier spectrum $\rho(\mathbf{q},-E_{1})$
(b). The density of states is measured with respect to its spatial
average value. The image window of $25 \times 25$ plaquettes is
taken from a system of size $1000 \times 1000$. Here the
impurity-induced Friedel oscillation along the diagonals have been
filtered away; see main text for details.} \label{FIG:IMAGE}
\end{figure}

To explore this spatial variation of the LDOS in more detail, we
have calculated the LDOS, with the energy fixed at $-E_1$, in the
vicinity of the impurity with and without the mode coupling. In
the absence of mode coupling, we obtain the results (not shown)
which are consistent with previous studies~\cite{Balatsky95}: the
LDOS exhibits a Friedel oscillation along the diagonals of
CuO$_{\rm 2}$ plane but no other non-trivial features. When the
mode coupling is turned on, in addition to the Friedel oscillation
along the diagonals, the LDOS displays a new type of modulation
with a period of $2a$ in the wide regions along the bond
direction.  This new modulation can be more easily seen
Fig.~\ref{FIG:IMAGE}(a) when the pre-dominant Friedel oscillation
is filtered away. It occurs in four disconnected triangles in the
field of view.

Also in order to highlight the new modulation, we find it useful
to perform a filtered Fourier transform, $\rho({\bf q},E) =
\sum_{i}^{\prime} {\rm e}^{i {\bf q} \cdot \mathbf{r}_{i}}
\rho(\mathbf{r}_{i},E)$, where $\sum_{i}^{\prime}$ denotes a
summation over all the sites in four triangles in
Fig.~\ref{FIG:IMAGE}(a).
The resulting Fourier-transformed image~\cite{note3} is given in
Fig.~\ref{FIG:IMAGE}(b), which unambiguously shows that the new
feature induced by the coupling to the spin resonance mode has a
spatial modulation wavevector $(\pi,\pi)$.


This new type of LDOS modulation with a wavevector close to
${\bf Q} = (\pi,\pi)$ reflects the dominance of the collective mode
effect near the $M$ points of the Brillouin zone.
It follows from Eqs.~(\ref{gij},\ref{ldos}),
with the form of the $T$-matrix under
consideration and
when the chosen field of view
has an inversion symmetry with respect to the
impurity site,
that the
Fourier transformed LDOS for
any finite ${\bf q}$ is,
\begin{eqnarray}
\rho({\bf q}, E)=-\frac{2}{\pi} \mbox{Im} \int d {\bf p}
\left [ {\underline{G}}_{0}({\bf p}+\mathbf{q};E) \hat{T}(E)
{\underline{G}}_{0}({\bf p},E) \right ]_{11}\;.
\label{convolution}
\end{eqnarray}
At positive energies, the poles shown in Fig.~\ref{FIG:SPECTRUM}
for ${\underline{G}}_{0}({\bf p};E)$ are
located~\cite{Dessau91,Ding96,Abanov02,Kee02,Eschrig00} at $E_{\bf
p}$ and $E_{{\bf p}-{\bf Q}} + \Omega_0$, respectively. (Here,
$E_{\bf p} \equiv \sqrt{\xi_{\bf p}^2 + \Delta_{\bf p}^2}$.)
Likewise, ${\underline{G}}_{0}({\bf p}+{\bf q};E)$ is the sum of
two poles, one at $E_{{\bf p}+{\bf q}}$ and the other at $E_{{\bf
p}+{\bf q}-{\bf Q}} + \Omega_0$. At $E=E_1$,
Eq.~(\ref{convolution}) is dominated by the convolution between
the poles at $E_{{\bf p}-{\bf Q}} + \Omega_0$ and $E_{{\bf p}+{\bf
q}-{\bf Q}} + \Omega_0$. This term in turn is dominated by the
contributions corresponding to when both ${\bf p}-{\bf Q}$ and
${\bf p}+{\bf q}-{\bf Q}$ are near to the $M$ points, leading to a
$\rho({\bf q}, -E_1)$ that is peaked near $({\pi, \pi})$.

Similar phase-phase considerations show that the vertex correction
terms lead to a similar momentum dependence. In the presence of
vertex corrections, the $T$-matrix satisfies an integral equation
in the wavevector space. The vertex correction to the $T$-matrix
to the same order ($g^2$) of our  calculation for the self-energy
is shown in Fig.~\ref{FIG:diagrams}(a). It involves a wavevector
convolution of a form similar to that given in
Eq.~(\ref{convolution}). We therefore expect~\cite{note2} an
additive contribution to $\rho({\bf q}, -E_1)$ that is also peaked
near $( {\pi, \pi})$.

Finally, we remark on issues which go beyond the idealized model
we have considered so far: (i) In the presence of a gap
inhomogeneity at the nanoscale, the mode signature would appear at
$\langle\Delta_{0}\rangle+\omega_0$, with a slight smearing. Here,
$\langle \Delta_{0} \rangle $ is the spatially averaged
superconducting gap. The fact that a well-defined peak-dip-hump
structure appears in the
break-junction tunneling spectrum~\cite{Zasadzinski01} implies
that the smearing is not too large; (ii) The mode signature
depends on the detailed band structure. In the cuprates, the band
is flat near the antinodal points (i.e., $(\pi,0)$ {\it etc.}),
and a mode with a momentum significantly different from
$(\pi,\pi)$ will have a weaker effect compared to that of the
$(\pi,\pi)$ mode we addressed; (iii)
Tunneling matrix elements need to be taken into account in order to
understand the detailed spatial variation of the LDOS
as observed around zinc impurities in BSCCO~\cite{Pan00,Martin02}.
Such a filtering effect, however, will not affect our
conclusion on the momentum of the LDOS modulation.

To summarize, we have studied the effects of the magnetic
resonance mode on the tunneling spectrum in the presence of a
nonmagnetic impurity. The LDOS around the impurity displays
resonant features at relatively high energies [close to $\pm
(\Omega_0+\Delta_0)$], which modulates in space with a wavevector
close to $(\pi,\pi)$. Our prediction can be tested
straightforwardly by operating the existing high resolution STM at
a relatively high energy window. Such studies should shed
considerable new light on the physics of the spin resonance mode,
in particular its coupling to the electronic excitations.

We are especially grateful to J. C. Davis for helpful
conversations at the early stage of this work. The authors have
benefited considerably from discussions with Ar. Abanov, Y. Bang,
A. V. Chubukov, K. Damle, and M. Norman. This work was supported
by the US DOE (JXZ and AVB), by TcSAM and the NSF Grant No.
DMR-0090071 (JS and QS). JXZ also acknowledges the hospitality of
the Rice University, where part of this research was carried out.


\begin{thebibliography}{99}
\bibitem{note3} Similar procedure for the LDOS in the absence of
a mode coupling yields a Fourier-transformed LDOS that is featureless
near $(\pi,\pi)$.
\bibitem{note2} We have numerically calculated the leading
order vertex correction contribution to the LDOS, finding that it
is indeed peaked near $(\pi,\pi)$.
\end{thebibliography}
\end{document}